\documentclass[12pt]{article}
\usepackage{fancyhdr}

\usepackage{mathrsfs}
\usepackage[T1]{fontenc}
\usepackage{mathpazo}
\usepackage{setspace}
\usepackage{amsfonts}
\usepackage{amssymb}
\usepackage{amsmath}
\usepackage{epsfig}
\usepackage{latexsym}
\usepackage{color}
\usepackage{graphicx}
\usepackage{nicefrac}
\usepackage[latin1]{inputenc}
\usepackage{slashed}
\usepackage{multirow}
\usepackage{fancybox}
\usepackage{cite}
\usepackage{comment}
\usepackage{soul}

\usepackage[all,cmtip]{xy}
\usepackage{hyperref}

\def\hybrid{\topmargin -20pt    \oddsidemargin 0pt
        \headheight 0pt \headsep 0pt
        \textwidth 6.25in       
        \textheight 9.25in       
        \marginparwidth .875in
        \parskip 5pt plus 1pt   \jot = 1.5ex}

\hybrid

\def\baselinestretch{1.2}

\catcode`\@=11

\def\marginnote#1{}
\newcount\hour
\newcount\minute
\newtoks\amorpm
\hour=\time\divide\hour by60
\minute=\time{\multiply\hour by60 \global\advance\minute by-\hour}
\edef\standardtime{{\ifnum\hour<12 \global\amorpm={am}%
        \else\global\amorpm={pm}\advance\hour by-12 \fi
        \ifnum\hour=0 \hour=12 \fi
        \number\hour:\ifnum\minute<10 0\fi\number\minute\the\amorpm}}
\edef\militarytime{\number\hour:\ifnum\minute<10 0\fi\number\minute}

\def\draftlabel#1{{\@bsphack\if@filesw {\let\thepage\relax
   \xdef\@gtempa{\write\@auxout{\string
      \newlabel{#1}{{\@currentlabel}{\thepage}}}}}\@gtempa
   \if@nobreak \ifvmode\nobreak\fi\fi\fi\@esphack}
        \gdef\@eqnlabel{#1}}
\def\@eqnlabel{}
\def\@vacuum{}
\def\draftmarginnote#1{\marginpar{\raggedright\scriptsize\tt#1}}

\def\draft{\oddsidemargin -.5truein
        \def\@oddfoot{\sl preliminary draft \hfil
        \text\thepage\hfil\sl\today\quad\militarytime}
        \let\@evenfoot\@oddfoot \overfullrule 3pt
        \let\label=\draftlabel
        \let\marginnote=\draftmarginnote
   \def\@eqnnum{(\theequation)\rlap{\kern\marginparsep\tt\@eqnlabel}%
\global\let\@eqnlabel\@vacuum}  }

\def\preprint{\twocolumn\sloppy\flushbottom\parindent 2em
        \leftmargini 2em\leftmarginv .5em\leftmarginvi .5em
        \oddsidemargin -.5in    \evensidemargin -.5in
        \columnsep .4in \footheight 0pt
        \textwidth 10.in        \topmargin  -.4in
        \headheight 12pt \topskip .4in
        \textheight 6.9in \footskip 0pt
        \def\@oddhead{\thepage\hfil\addtocounter{page}{1}\thepage}
        \let\@evenhead\@oddhead \def\@oddfoot{} \def\@evenfoot{} }

\def\numberbysection{\@addtoreset{equation}{section}
        \def\theequation{\thesection.\arabic{equation}}}

\def\underline#1{\relax\ifmmode\@@underline#1\else
        $\@@underline{\hbox{#1}}$\relax\fi}

\def\titlepage{\@restonecolfalse\if@twocolumn\@restonecoltrue\onecolumn
     \else \newpage \fi \thispagestyle{empty}\c@page\z@
        \def\thefootnote{\fnsymbol{footnote}} }

\def\endtitlepage{\if@restonecol\twocolumn \else \newpage \fi
        \def\thefootnote{\arabic{footnote}}
        \setcounter{footnote}{0}}  

\catcode`@=12
\relax

\def\figcap{\section*{Figure Captions\markboth
        {FIGURECAPTIONS}{FIGURECAPTIONS}}\list
        {Figure \arabic{enumi}:\hfill}{\settowidth\labelwidth{Figure
999:}
        \leftmargin\labelwidth
        \advance\leftmargin\labelsep\usecounter{enumi}}}
 \relax
\def\tablecap{\section*{Table Captions\markboth
        {TABLECAPTIONS}{TABLECAPTIONS}}\list
        {Table \arabic{enumi}:\hfill}{\settowidth\labelwidth{Table
999:}
        \leftmargin\labelwidth
        \advance\leftmargin\labelsep\usecounter{enumi}}}
 \relax
\def\reflist{\section*{References\markboth
        {REFLIST}{REFLIST}}\list
        {[\arabic{enumi}]\hfill}{\settowidth\labelwidth{[999]}
        \leftmargin\labelwidth
        \advance\leftmargin\labelsep\usecounter{enumi}}}
 \relax
\makeatletter
\newcounter{pubctr}
\def\publist{\@ifnextchar[{\@publist}{\@@publist}}
\def\@publist[#1]{\list
        {[\arabic{pubctr}]\hfill}{\settowidth\labelwidth{[999]}
        \leftmargin\labelwidth
        \advance\leftmargin\labelsep
        \@nmbrlisttrue\def\@listctr{pubctr}
        \setcounter{pubctr}{#1}\addtocounter{pubctr}{-1}}}
\def\@@publist{\list
        {[\arabic{pubctr}]\hfill}{\settowidth\labelwidth{[999]}
        \leftmargin\labelwidth
        \advance\leftmargin\labelsep
        \@nmbrlisttrue\def\@listctr{pubctr}}}
 \relax
\makeatother
\newskip\humongous \humongous=0pt plus 1000pt minus 1000pt

\newif\ifdtup

\relax

\def\be{\begin{equation}}
\def\ee{\end{equation}}
\def\ba{\begin{eqnarray}}
\def\ea{\end{eqnarray}}

\def\del{\partial}

\def\k{\kappa}

\def\b{\beta}

\def\d{\delta}

\def\m{\mu}
\def\n{\nu}

\def\l{\lambda}
\def\L{\Lambda}
\def\s{\sigma}

\def\no{\noindent}

\def\qq{\qquad}

\def\IR{\relax{\text I\kern-.18em R}}

\def \ha {{1\over 2}}

\def \ov {\over}

\def\IR{\relax{\text I\kern-.18em R}}
\def\IL{\relax{\text I\kern-.18em L}}

\def\inv{^{\raise.15ex\hbox{${\scriptscriptstyle -}$}\kern-.05em 1}}

\def\Tr{{\text Tr}}

\begin{document}

\renewcommand{\theequation}{\thesection.\arabic{equation}}
\csname @addtoreset\endcsname{equation}{section}

\newcommand{\beq}{\begin{equation}}
\newcommand{\eeq}[1]{\label{#1}\end{equation}}
\newcommand{\ber}{\begin{eqnarray}}
\newcommand{\eer}[1]{\label{#1}\end{eqnarray}}
\newcommand{\eqn}[1]{(\ref{#1})}
\begin{titlepage}
\begin{center}

~

\vskip  .3in

{\large\bf Double and cyclic $\lambda$-deformations and their canonical equivalents}

\vskip 0.35in

{\bf George Georgiou,$^1$\ \ Konstantinos Sfetsos}$^2$\ \ and\ \ {\bf Konstantinos Siampos}$^3$
\vskip 0.1in

\vskip 0.1in
{\em
${}^1$Institute of Nuclear and Particle Physics,\\ National Center for Scientific Research Demokritos,\\
Ag. Paraskevi, GR-15310 Athens, Greece
}
\vskip 0.1in

 {\em
${}^2$Department of Nuclear and Particle Physics,\\
Faculty of Physics, National and Kapodistrian University of Athens,\\
Athens 15784, Greece\\
}
\vskip 0.1in

{\em${}^3$Albert Einstein Center for Fundamental Physics,\\
Institute for Theoretical Physics / Laboratory for High-Energy Physics,\\
University of Bern,
Sidlerstrasse 5, CH3012 Bern, Switzerland
}

\vskip 0.1in

{\footnotesize \texttt georgiou@inp.demokritos.gr, ksfetsos@phys.uoa.gr, siampos@itp.unibe.ch}


\end{center}

\vskip 0.1in

\centerline{\bf Abstract}
\no
We prove that the doubly $\lambda$-deformed $\sigma$-models, which include integrable cases,
are canonically equivalent to the sum of two single $\lambda$-deformed models.
This explains the equality of the exact $\beta$-functions and current anomalous dimensions
of the doubly $\lambda$-deformed $\sigma$-models to those of two single $\lambda$-deformed models.
Our proof is based upon agreement of their Hamiltonian densities and of their canonical structure. Subsequently,
we show that it is possible to take a well defined non-Abelian type limit of the doubly-deformed action.
Last, but not least, by extending the above, we construct multi-matrix integrable deformations of an arbitrary number of
 WZW models.

\newpage

\end{titlepage}

\tableofcontents


\def\baselinestretch{1.2}
\baselineskip 20 pt
\noindent


\setcounter{equation}{0}
\renewcommand{\theequation}{\thesection.\arabic{equation}}

\section*{Introduction and results}
\addcontentsline{toc}{section}{Introduction and results}

A new class of integrable theories based on current algebras for a semi-simple group was recently constructed 
\cite{Georgiou:2016urf}.
The starting point was to consider two independent WZW models
at the same positive integer level $k$ and two distinct PCM models which were then
left-right asymmetrically gauged with respect to a common global symmetry.
The models are labeled by the level $k$ and two
general invertible matrices $\lambda_{1,2}$. For certain choices of $\lambda_{1,2}$ integrability
is retained \cite{Georgiou:2016urf}.
This idea can be generalized to include integrable deformations of exact CFTs on symmetric spaces.
This construction is reminiscent to the one for single $\lambda$-deformations 
\cite{Sfetsos:2013wia,Hollowood:2014rla,Hollowood:2014qma}.

\no
Subsequently, the quantum properties of the aforementioned multi-parameter integrable deformations were studied in
\cite{Georgiou:2017aei}, by employing a variety of techniques. One of the main results of that work was
that the running of the
couplings $\lambda_1$ and $\lambda_2$, as well as the anomalous dimensions of current operators depend only on one of the
couplings, either
 $\lambda_1$ or $\lambda_2$ and are identical to those found for single $\lambda$-deformations
\cite{Kutasov:1989dt,Gerganov:2000mt,Itsios:2014lca, Sfetsos:2014jfa,
Georgiou:2015nka,Georgiou:2016iom,Georgiou:2016zyo}.
These rather unexpected results seek for a simple explanation. The purpose of this work is to
demonstrate that they are due to the fact that the doubly deformed models are
canonically equivalent to the sum of two single $\lambda$-deformations, one with deformation matrix
being $\lambda_1$ and the other with deformation matrix $\lambda_2$.
Recall that all known forms of T-duality, i.e., Abelian, non-Abelian and Poisson--Lie T-duality
can be formulated as canonical transformations in the phase space of the corresponding two-dimensional $\sigma$-models
\cite{Curtright:1994be,Alvarez:1994wj,Lozano:1995jx,Sfetsos:1996xj,Sfetsos:1997pi}.
Moreover, it has been shown in various works that
the running of couplings is preserved under these canonical transformations even though the corresponding 
$\sigma$-models fields
are totally different \cite{Balazs:1997be,Balog:1998br,Sfetsos:1998kr, Sfetsos:1999zm,Sfetsos:2009dj}. All of the
above strongly hint towards the validity of our assertion, which of course we will prove.

The plan of the paper is as follows: In section \ref{canonicalsection}, after a brief review
of the single and doubly $\lambda$-deformed models and of their {\it non-perturbative} symmetries, we will
show that the doubly deformed models are canonically equivalent to the sum of two single $\lambda$-deformations.
In section \ref{newnonabeliansection}, we will present the type of non-Abelian T-duality that
is based on the doubly deformed $\sigma$-models of \cite{Georgiou:2016urf}.
Finally, in section \ref{cycliclambdasection}, we will construct multi-matrix {\it integrable} deformations of an arbitrary
number of independent WZW models by performing a left-right asymmetric gauging for each one of them but in
such a way that the total classical gauge anomaly vanishes. This happens if these models are forced to obey the
cyclic symmetry property or if they are infinitely many, resembling
in structure either a closed or an infinitely open spin chain. Their action can be thought of as the all-loop
effective action of several independent WZW models for $G$ all at level $k$, perturbed by current bilinears
mixing the different WZW models with nearest neighbour-type interactions. These models are also canonically
equivalent to a sum of single $\lambda$-deformed models with appropriate couplings.
Furthermore, we will argue that the Hamiltonian of these new models maps to itself under an inversion
of all couplings $\lambda_i \mapsto \lambda_i^{-1}, \,\,i=1,...,n$ accompanied generically by non-local redefinitions of the
group elements involved when $n=3,4,\dots$. This symmetry, which in the special cases where $n=1,2$ simplifies to the one
reviewed in section \ref{canonicalsection}, is in accordance with the fact that the $\b$-functions and
anomalous dimensions of currents are again given by the same expressions as in the case of the single 
$\lambda$-deformed model.

\section{Review and canonical equivalence}
\label{canonicalsection}
\subsection{Single $\lambda$-deformed $\s$-models}
\label{singlelambdalkslsl}

{
The construction of the single $\lambda$-deformed $\sigma$-model starts by considering 
the sum of a gauged WZW and a PCM for a group $G$, defined with group elements $g$ and $\tilde g$, respectively and next gauging the global symmetry \cite{Sfetsos:2013wia} 
\begin{equation*}
g\mapsto\Lambda^{-1}g\Lambda\,,\quad \tilde g\mapsto\Lambda^{-1}\tilde g\ .
\end{equation*}
This is done by introducing gauge fields $A_\pm$ in the Lie-algebra of $G$ transforming as
\begin{equation*}
A_\pm\mapsto\Lambda^{-1}A_\pm\Lambda-\partial_\pm\Lambda\,.
\end{equation*}
The choice $\tilde g=\mathbb{I}$ completely fixes the gauge and the gauged fixed action reads 
\be
\label{kdldjkdfkj}
\begin{split}
&
S_{k,\lambda}(g_;A_{\pm}) =S_k(g)
+ {k\ov \pi} \int \text{d}^2\s\, \Tr\Big(
A_- \del_+ g g^{-1} - A_+ g^{-1} \del_- g
\\
&
\qq\qq \qq + A_- g A_+\, g^{-1} -A_+ \lambda^{-1} A_-\Big)\ ,
\end{split}
\ee
where $S_k(g)$ is the WZW model.
The $A_{\pm}$'s are non-dynamical and their equations of motion read
\be
\label{kshdg1}
\nabla_+ g\, g^{-1} =  (\l^{-T}-\mathbb{I}) A_+ \ ,\quad
g^{-1} \nabla_- g = - (\l^{-1}-\mathbb{I}) A_-\,,
\ee
with $\nabla_\pm g=\partial_\pm g-[A_\pm,g]$.
Solving them in terms of the gauge fields we find
\begin{equation}
\label{sklldldd}
A_+=i\left(\l^{-T}-D\right)^{-1}J_+\,,\quad A_-=-i\left(\l^{-1}-D^T\right)^{-1}J_-\,,
\end{equation}
where
\be
\label{hg3}
J^a_+ = - i\, \Tr(t_a \del_+ g g^{-1}) ,\qq J^a_- = - i\, \Tr(t_a g^{-1}\del_- g )\ .
\qq D_{ab}= \Tr(t_a g t_b g^{-1})\ ,
\ee
where $t_a$'s are Hermitian representation matrices obeying $[t_a,t_b]=if_{abc}t_c$, so that
the structure constants $f_{abc}$ are real. We choose the normalization such that $\Tr(t_at_b)=\d_{ab}$.

\no
Using \eqref{sklldldd} into \eqref{kdldjkdfkj} one finds the action \cite{Sfetsos:2013wia}
\be
S_{k,\l}(g)=  S_k(g)+ {k\ov \pi} \int \text{d}^2\s\,{\rm Tr}\left(J_+(\l^{-1}- D^T)^{-1} J_-\right)\ .
\label{laact}
\ee
For small elements of the matrix $\lambda$ this action becomes
\begin{equation*}
S_{k,\l}(g) = S_k(g) + {k\ov \pi}
\int \text{d}^2\s\, {\rm Tr}\left(J_{+}\l J_{-}\right)  + \cdots \ .
\end{equation*}
Hence \eqref{laact} represents the effective action of self-interacting current bilinears of a single WZW model.
The action \eqref{laact} has the remarkable {\it non-perturbative} symmetry \cite{Itsios:2014lca,Sfetsos:2014jfa}
\be
k\mapsto -k \ ,\quad \l \mapsto  \l^{-1}\ , \quad g\mapsto g^{-1}  \ .
\label{duallli}
\ee
As in the case of gauged WZW models \cite{Bowcock:1988xr}, we define the currents $\mathcal{J}_\pm$
\be
\begin{split}
\label{singleAffine}
\mathcal{J}_{+} =\nabla_+ g g^{-1} + A_{+} - A_{-}\ ,\quad \mathcal{J}_-  = - g^{-1} \nabla_- g+ A_{-} - A_{+}\ ,
\end{split}
\ee
The above form for the $\mathcal{J}_{\pm}^a$'s when rewritten in terms of phase space
variables of the $\sigma$-model action, assumes the same form as the currents $J_\pm^a$ of the WZW action.
Hence, they satisfy two commuting current algebras as in \cite{Witten:1983ar}
\be
\label{singleBowcock}
\{\mathcal{J}_{\pm}^a , \mathcal{J}_{\pm}^b\} =
\frac{2}{k}\,f_{abc} \mathcal{J}_{\pm}^c \d_{\s\s'} \pm \frac{2}{k} \d_{ab} \d'_{\s\s'}\,,\quad
\d_{\s\s'}=\d(\s-\s')\ .
\ee
Moreover using \eqref{kshdg1} we can rewrite \eqref{singleAffine} as
\be
\label{singleSthroughA}
\mathcal{J}_+ =\l^{-T} A_{+} - A_{-}\ ,\quad \mathcal{J}_- = \l^{-1} A_{-} - A_{+}\ .
\ee
Inversely
\be
\label{singleAthroughS}
\begin{split}
&A_{+} = h^{-1} \l^T (\mathcal{J}_+ + \l \mathcal{J}_-)\ ,
\quad A_{-} =  \tilde h^{-1} \l (\mathcal{J}_- + \l^T \mathcal{J}_+)\ ,\\
&h=\mathbb{I}-\l^T\l\,,\quad \tilde h=\mathbb{I}-\l\l^T\,,
\end{split}
\ee
assuming that the matrix $\l$ is such that $h,\tilde h$ are positive-definite matrices. To obtain the Poisson algebra in the base of $A_\pm$
we use \eqref{singleBowcock}, \eqref{singleSthroughA} and \eqref{singleAthroughS}. 

To study the Hamiltonian structure of the problem we need to define its phase space \cite{Hollowood:2014rla,Hollowood:2014qma}.
This is given in terms of the currents $\mathcal{J}_{\pm}$, the gauge fields $A_\pm$
and the associated momenta $P_\pm$ to $A_\pm$. The $\mathcal{J}_{\pm}$
obey two commuting current algebras \eqref{singleBowcock}
and have vanishing Poisson brackets with $A_\pm$ and $P_\pm$
\begin{equation*}
\{P_\pm^a(\s),A^b_\mp(\s')\}=\d^{ab}\d(\s-\s')\,.
\end{equation*}
Furthermore, since the $A_{\pm}$'s are non-dynamical their associated momenta $P_\pm$ vanish.
This introduces two primary constraints
\begin{equation*}
\varphi_1=P_+\approx0\,,\quad \varphi_2=P_-\approx0\, .
\end{equation*}
Their time-evolution gives rise to the secondary constraints
\begin{equation*}
\varphi_3=\mathcal{J}_+ -\l^{-T} A_{+} + A_{-}\approx0\,,\quad
\varphi_4=\mathcal{J}_- - \l^{-1} A_{-} +A_{+}\approx0\ .
\end{equation*}
Time evolution generates no further constraints.
The $\varphi_i$'s with $i=1,2,3,4$, turn out to be second class constraints, since the matrix of their Poisson brackets is invertible in the deformed case.
Finally, the Hamiltonian density of the single $\lambda$-deformed model before integrating out the gauge fields takes
the form \cite{Sfetsos:2013wia,Bowcock:1988xr}
\begin{equation*}
\begin{split}
\mathcal{H}_{\text{single}}&=\frac{k}{4\pi}{\rm Tr}\left(\mathcal{J}_+\mathcal{J}_++\mathcal{J}_-\mathcal{J}_-\right.
\left. +4(\mathcal{J}_+A_{-}+\mathcal{J}_-A_{+})\right.\\
&\left. +2(A_{+}-A_{-})(A_{+}-A_{-})\right.
\left.-4A_{+}(\lambda_1^{-1}-\mathbb{I}) A_{-}\right),
\end{split}
\end{equation*}
or equivalently through \eqref{singleSthroughA}, in terms of $A_{\pm}$'s
\be
\boxed{
\label{singleHdensity0}
\mathcal{H}_{\text{single}}=\frac{k}{4\pi}\,{\rm Tr}\left(A_{+}\left(\l^{-1}\tilde h\l^{-T}\right)A_{+}
+A_{-}\left(\l^{-T} h\l^{-1}\right)A_{-}\right) } \,.
\ee

}

\subsection{Doubly $\lambda$-deformed $\s$-models}
\label{doublyldeformed}

 The action defining the doubly deformed models depends on two group elements $g_i\in G,\,\,\,i=1,2$ and is given
by the deformation of the sum of two WZW models $S_k(g_1)$ and $S_k(g_2)$ as \cite{Georgiou:2016urf}
\be
\begin{split}
&  S_{k,\lambda_1,\lambda_2}(g_1,g_2) = S_k(g_1) + S_k(g_2)
\\
&\qq\quad + {k\ov \pi} \int  \text{d}^2\s\, {\rm Tr}\left\{ \left(\!\! \begin{array}{cc}
    J_{1+}\! &\! J_{2+} \end{array}\!\!  \right)
\left(  \begin{array}{cc}
     \L_{21}\lambda_1 D_2^T\lambda_2 &   \L_{21}\lambda_1 \\
     \L_{12}\lambda_2 & \L_{12} \lambda_2 D_1^T\lambda_1\\
  \end{array} \right)
  \left(\!\! \begin{array}{c}
    J_{1-} \\ J_{2-} \end{array}\!\!\! \right)\right\} \ ,
\end{split}
\label{defactigen}
\ee
where
\begin{equation}
\label{doubleLambdas}
\L_{12}= (\mathbb{I} - \lambda_2 D_1^T \lambda_1 D_2^T)^{-1}\ ,\quad
\L_{21}= (\mathbb{I} - \lambda_1 D_2^T \lambda_2 D_1^T)^{-1}\ .
\end{equation}
The matrices $D_{ab}$ and the currents $J^a_{\pm}$ are defined in  \eqref{hg3}.
When a current or the matrix $D$ has the extra index $1$ or $2$ this means that one should use the
corresponding group element in its definition.
The action \eqref{defactigen} has the {\it non-perturbative} symmetry \cite{Georgiou:2016urf}
\be
 k \mapsto -k \ , \quad \lambda_1 \mapsto  \lambda_1^{-1} \ ,\quad  \lambda_2 \mapsto  \lambda_2^{-1}\ ,
 \quad g_1\mapsto g_2^{-1}\,,\quad g_2\mapsto g_1^{-1}\ ,
\label{symmdual}
\ee
which is similar to \eqref{duallli}.
For small elements of the matrices $\lambda_i$'s the action \eqref{defactigen} becomes
\begin{equation*}
S_{k,\lambda_1,\lambda_2}(g_1,g_2) = S_k(g_1) + S_k(g_2) + {k\ov \pi}
\int \text{d}^2\s\ {\rm Tr}(J_{1+}\lambda_1 J_{2-} + J_{2+}\lambda_2 J_{1-}) + \cdots\ .
\end{equation*}
Hence \eqref{defactigen} represents the effective action of two WZW models mutually interacting via current bilinears.
Similarly to \eqref{singleAffine}
we define the currents\footnote{To conform with
the notation of the current work we have renamed the gauged fields
$(A_\pm,B_\pm)$ of \cite{Georgiou:2016urf} by $(A^{(1)}_{\pm},A^{(2)}_{\pm})$.}\textsuperscript{,}\footnote{
The various covariant derivatives are defined according to the transformation properties of
the object they act on. For instance
\begin{equation*}
\nabla_\pm g_1= \del_\pm g_1 -A^{(1)}_{\pm} g_i + g_iA^{(2)}_{\pm}\,,\quad
\nabla_\pm (\nabla_\mp g_1  g_1^{-1})=
\del_\pm(\nabla_\mp g_1  g_1^{-1}) -[A^{(1)}_\pm,\nabla_\mp g_1  g_1^{-1}]\,.
\end{equation*}
}
\be
\begin{split}
\label{Affine}
&\mathcal{J}^{(1)}_{+}  =\nabla_+ g_1 g_1^{-1} + A^{(1)}_+ - A^{(1)}_-\ ,
\quad \mathcal{J}^{(1)}_{-}  = - g_1^{-1} \nabla_- g_1+ A^{(2)}_{-} - A^{(2)}_{+}\ ,\\
&\mathcal{J}^{(2)}_{+}  =\nabla_+ g_2 g_2^{-1} + A^{(2)}_{+} - A^{(2)}_{-}\ ,
\quad \mathcal{J}^{(2)}_{-}  =- g_2^{-1} \nabla_- g_2 + A^{(1)}_- - A^{(1)}_+ \ .
\end{split}
\ee
These currents obey two commuting copies of current algebras \cite{Georgiou:2016urf}
\be
\label{Bowcock}
\{\mathcal{J}_{\pm}^{(i)a} , \mathcal{J}_{\pm}^{(i)b}\} =
\frac{2}{k}\,f_{abc} \mathcal{J}_{\pm}^{(i)c} \d_{\s\s'} \pm \frac{2}{k} \d_{ab} \d'_{\s\s'}\,,\quad i=1,2\ ,
\ee
which encode the canonical structure of the theory.
The action does not depend on derivatives of $A^{(i)}_\pm\,, i=1,2$, so that as in subsection \ref{singlelambdalkslsl},
their equations of motion
are second class constraints \cite{Georgiou:2016urf}
\be
\label{kshdg2}
\begin{split}
&\nabla_+ g_1\, g_1^{-1} =  (\lambda_1^{-T}-\mathbb{I}) A^{(1)}_+ \ ,\quad
g_1^{-1} \nabla_- g_1 = - (\lambda_2^{-1}-\mathbb{I}) A^{(2)}_{-}\,,\\
&\nabla_+ g_2\, g_2^{-1} =  (\lambda_2^{-T}-\mathbb{I}) A^{(2)}_{+} \ ,\quad
g_2^{-1} \nabla_- g_2 = - (\lambda_1^{-1}-\mathbb{I}) A^{(1)}_-\,,
\end{split}
\ee
determining the gauge fields in terms of the group elements similarly to \eqn{sklldldd} (for
the precise expressions we refer to \cite{Georgiou:2016urf}). Then \eqref{Affine} rewrites as
\be
\label{SthroughA}
\begin{split}
&\mathcal{J}^{(1)}_{+} = \lambda_1^{-T} A^{(1)}_{+} - A^{(1)}_{-}\ ,
\quad \mathcal{J}^{(1)}_{-} = \lambda_2^{-1} A^{(2)}_- - A^{(2)}_+\ ,\\
&\mathcal{J}^{(2)}_{+} = \lambda_2^{-T} A^{(2)}_{+} - A^{(2)}_{-}\ ,
\quad \mathcal{J}^{(2)}_{-} = \lambda_1^{-1} A^{(1)}_- - A^{(1)}_+\ .
\end{split}
\ee
Equivalently the gauge fields in terms of the dressed currents are given by
\be
\label{AthroughS}
\begin{split}
&A^{(1)}_+ = h_1^{-1} \lambda_1^T (\mathcal{J}^{(1)}_{+} + \lambda_1\, \mathcal{J}^{(2)}_{-})\ ,
\quad A^{(1)}_- =  \tilde h_1^{-1} \lambda_1 (\mathcal{J}^{(2)}_{-} + \lambda_1^T\, \mathcal{J}^{(1)}_{+})\ ,\\
&A^{(2)}_{+} =  h_2^{-1} \lambda_2^T (\mathcal{J}^{(2)}_{+} + \lambda_2\, \mathcal{J}^{(1)}_{-})\ ,
\quad A^{(2)}_{-} =   \tilde h_2^{-1} \lambda_2 (\mathcal{J}^{(1)}_{-} + \lambda_2^T\, \mathcal{J}^{(2)}_{+})\ ,\\
&h_i=\mathbb{I}-\lambda_i^T\lambda_i\,,\quad \tilde h_i=\mathbb{I}-\lambda_i\lambda_i^T\,,\quad i=1,2\,.
\end{split}
\ee
To obtain the Poisson algebra in the base of $A^{(1)}_{\pm}$ and $A^{(2)}_{\pm}$
we use \eqref{Bowcock}, \eqref{SthroughA} and \eqref{AthroughS}.
As a corollary one can easily show that
$\{A^{(1)}_\pm,A^{(2)}_\pm\}=0$\,,
for all choices of signs and for generic coupling matrices $\lambda_{1,2}$.
The Hamiltonian density of our system before integrating out the gauge fields takes the form \cite{Georgiou:2016urf}
\begin{equation*}
\begin{split}
\mathcal{H}_{\text{doubly}}&=
\frac{k}{4\pi}{\rm Tr}\left\{\mathcal{J}^{(1)}_{+}\mathcal{J}^{(1)}_{+}
+\mathcal{J}^{(1)}_{-}\mathcal{J}^{(1)}_{-}+\mathcal{J}^{(2)}_{+}\mathcal{J}^{(2)}_{+}+
\mathcal{J}^{(2)}_{-}\mathcal{J}^{(2)}_{-}\right.\\
&\left. +4(\mathcal{J}^{(1)}_{+}A^{(1)}_{-}+\mathcal{J}^{(2)}_{+}A^{(2)}_{-}+\mathcal{J}^{(1)}_{-}A^{(2)}_{+}+
\mathcal{J}^{(2)}_{-}A^{(1)}_{+})\right.\\
&\left. +2(A^{(1)}_{+}-A^{(1)}_{-})(A^{(1)}_{+}-A^{(1)}_{-})+2(A^{(2)}_{+}-A^{(2)}_{-})(A^{(2)}_{+}-A^{(2)}_{-})\right.\\
&\left.-4A^{(1)}_{+}(\lambda_1^{-1}-\mathbb{I})A^{(1)}_{-}-4A^{(2)}_{+}(\lambda_2^{-1}-\mathbb{I})A^{(2)}_{-}\right\}\
\end{split}
\end{equation*}
and can be rewritten through \eqref{SthroughA} in terms of $A^{(i)}_{\pm}$ and $\lambda_i$ as
\be
\label{Hdensity0}
\boxed{
\mathcal{H}_{\text{doubly}}=\frac{k}{4\pi}\sum_{i=1}^2{\rm Tr}\left(A_+^{(i)}\left(\lambda_i^{-1}\tilde h_i\lambda_i^{-T}\right)A_+^{(i)}+
A_-^{(i)}\left(\lambda_i^{-T} h_i\lambda_i^{-1}\right)A_-^{(i)}\right)}\,.
\ee

The fact that the Hamiltonian density \eqref{Hdensity0}  is the sum of two terms one depending on $A^{(1)}_{\pm}$
and the other on $A^{(2)}_{\pm}$  combined with the fact that the currents $\mathcal{J}^{(i)}_{\pm}, \,\,i=1,2,$
obey two commuting copies
of the current algebra of the single $\lambda$-deformed model shows that the doubly deformed models are canonically
equivalent to the sum of two single $\lambda$-deformed models, one with coupling $\lambda_1$ and the other with coupling
$\lambda_2$. The relations defining the canonical transformation are given by
\begin{equation}\label{can-def}
A^{(1)}_{\pm} =\tilde A^{(1)}_{\pm} \ ,\quad A^{(2)}_{\pm}=\tilde A^{(2)}_{\pm} \ ,
\end{equation}
where the gauge fields without the tildes correspond to the doubly deformed models and depend on
$(\lambda_1,\lambda_2;g_1,g_2)$, while the tilded gauge fields correspond to the canonically
equivalent sum of two single $\lambda$-deformed models the first of which depends on $(\lambda_1;\tilde g_1)$
only while the second depends on $(\lambda_2;\tilde g_2)$. 

Furthermore, the gauge fields of \eqref{can-def} should be considered as
functions of the coordinates parametrising the group elements and their conjugate momenta.
We may write relations involving world-sheet derivatives of the various group elements by using
\eqref{sklldldd} and \eqref{kshdg2}. As in all canonical transformation involving
canonical variables as well as their momenta, the relation between the $g_i$'s and the $\tilde g_i$'s is a non-local one.

A comment is in order concerning the $\eta$-deformed models
\cite{Klimcik:2002zj,Klimcik:2008eq,Delduc:2013fga,Delduc:2013qra,Arutyunov:2013ega}
which are closely related to the single $\lambda$-deformed ones via Poisson--Lie T-duality \cite{KS95a}
and an appropriate analytic continuation of the coordinates and the parameters
\cite{Klimcik:2015gba,Klimcik:2016rov,Vicedo:2015pna,Hoare:2015gda,Sfetsos:2015nya}
\begin{equation*}
{
\lambda\mapsto \frac{i E-\eta\mathbb{I}}{i E+\eta\mathbb{I}}\ ,
}
\end{equation*}
where $E$ is an arbitrary constant matrix.
Poisson--Lie T-duality can also be formulated as a canonical transformation \cite{Sfetsos:1996xj,Sfetsos:1997pi}
and therefore there is a chain of canonical transformations from doubly $\lambda$-deformed, to
two single $\lambda$-deformed and to $\eta$-deformed models.
It would be interesting to formulate the canonical transformation \eqref{can-def} via a duality invariant action similarly perhaps to the case of Poisson--Lie T-duality in \cite{Klimcik:1995dy}.

There is an important observation for further use in section \ref{cycliclambdasection}.
The Hamiltonian density \eqref{Hdensity0}
has the following {\it non-perturbative} symmetry
\be\label{npsym}
k \mapsto -k, \quad \lambda_i \mapsto \l^{-1}_i, \quad A^{(i)}_+ \mapsto \l^{-T}_i A^{(i)}_+,
\quad A^{(i)}_- \mapsto \l^{-1}_i A^{(i)}_-,\quad i=1,2.
\ee
In other words $\mathcal{H}_{\text{doubly}}$ maps to itself under \eqref{npsym}.
By using \eqref{SthroughA} this implies the following transformation for the group elements $g_1$ and $g_2$
\be\label{npsym-curr}
\mathcal{J}^{(1)}_{+} \mapsto -\mathcal{J}^{(2)}_{-} \,, \quad \mathcal{J}^{(2)}_{+} \mapsto -\mathcal{J}^{(1)}_{-} \,,
\quad \mathcal{J}^{(1)}_{-} \mapsto -\mathcal{J}^{(2)}_{+}\,, \quad \mathcal{J}^{(2)}_{-} \mapsto -\mathcal{J}^{(1)}_{+}\,.
\ee
Since the currents $\mathcal{J}^{(i)}_{\pm}, \,\,i=1,2$, depend both on the group elements and their derivatives, the
transformation \eqref{npsym-curr} can be viewed as a non-local transformation at the level of the group elements.
In the special cases of the single and doubly $\lambda$-deformed theories the symmetry \eqref{npsym}
and \eqref{npsym-curr} can be realized locally simply by a mapping of group elements,
i.e. \eqref{duallli} and \eqref{symmdual}. Indeed, it
is not difficult to check that \eqref{duallli} and \eqref{symmdual} imply for the gauge fields the transformation
\eqref{npsym}. The situation is slightly different for the generic cyclic models constructed below in
section \ref{cycliclambdasection} which can have arbitrarily many group elements.

\section{Doubly-deformed models and non-Abelian T-duality}
\label{newnonabeliansection}

It is has been known that the action \eqref{laact} admits the
non-Abelian T-dual limit that involves taking $k\to \infty$, whereas simultaneously taking the matrix $\lambda$ and the group
element $g$ to the identity \cite{Sfetsos:2013wia}.
Specifically, if we let
\begin{equation*}
\l= \mathbb{I} - {E\ov k}\ ,\qq g=\mathbb{I} + i {v\ov k}\ ,\qq k\to \infty\ ,
\end{equation*}
where $E$ is a constant matrix and $v=v_a t^a$, then the action \eqref{laact} becomes
\begin{equation*}
S(v,E)={1\ov \pi} \int \text{d}^2\s\, \Tr\left(\del_+ v (E+f)^{-1} \del_- v\right) \ ,
\end{equation*}
where $f$ is a matrix with elements $f_{ab}= f_{abc} v^c$.
This is the non-Abelian T-dual of the PCM action with general coupling matrix $E$
\begin{equation*}
S_{\rm PCM}(g,E)= -{1\ov \pi}\int \text{d}^2\s\, \Tr
\left(g^{-1}\del_+ g\, E\, g^{-1}\del_-g\right)\ ,
\end{equation*}
with respect to the global symmetry $g\mapsto \L\, g$, $\L\in G$.
The above limit is well defined when is taken on the $\b$-function for $\lambda$, as well as on the anomalous dimensions of various
operators in the theory. In the case of doubly $\lambda$ or even multiple/cyclic $\lambda$-deformations
(see section \ref{cycliclambdasection})
we have shown in particular that, the $\b$-functions and current anomalous dimensions are the same
with those of two or more simple $\lambda$-deformations.
Hence, it is expected that it should be possible to take a well defined non-Abelian type limit in the
action \eqref{defactigen}.
This is not necessarily simple since a suitable limit involves the two group elements.

In the following we focus on the most interesting case in which the matrices $\lambda_i$, $i=1,2$ are
isotropic, i.e. $(\lambda_i)_{ab}= \lambda_i\, \d_{ab}$.
It is convenient to use the group element
$\mathcal{G}=g_1 g_2$ and also rename $g_2$ by $g$. Then employing the  Polyakov--Wiegmann identity \cite{Polyakov:1983tt},
the action \eqref{defactigen}, using also \eqref{doubleLambdas}, takes the form
\ba
\label{rewr}
&&
S_{k,\lambda_1,\lambda_2}(\mathcal{G},g)= S_k(\mathcal{G}) +
{k\ov \pi} \int \text{d}^2\s\, \Tr\Big((1-\lambda_2)g^{-1}\del_+ g \left({\cal D}-\lambda_1\mathbb{I}\right)\Sigma\, g^{-1}\del_-g\\
&&
-(1-\lambda_2)g^{-1}\del_+ g\, \Sigma\,\del_-\mathcal{G} \mathcal{G}^{-1}
+ \lambda_1 (1-\lambda_2) \mathcal{G}^{-1}\del_+ \mathcal{G}\, \Sigma\, g^{-1}\del_-g
+ \lambda_1\lambda_2\, \mathcal{G}^{-1}\del_+\mathcal{G}\, \Sigma\, \mathcal{G}^{-1}\del_-\mathcal{G}\Big)\ ,\nonumber
\ea
where: $\Sigma=\left(\lambda_1\lambda_2 \mathbb{I} -{\cal D}\right)^{-1}$ and
 ${\cal D}=D(\mathcal{G})=D(g_1)D(g_2)$. Next we take the limit
\be
\label{dsh11}
\lambda_i=1-{\k_i^2\ov k}\ ,\quad i=1,2 \ ,\qquad \mathcal{G}= \mathbb{I} + i {v\ov k}\ ,\qq k\to \infty\ .
\ee
After some algebra we find that \eqref{rewr} becomes
\begin{equation}
\begin{split}
&
S_{\k_1^2,\k_2^2}(v,g) = -{1\ov \pi}\int \text{d}^2\s\, \Tr\Big( \k_2^2 g^{-1} \del_+ g g^{-1}\del_- g
\\
&\quad + \big(i \del_+ v  - \k_2^2 g^{-1} \del_+ g\big)\big((\k_1^2+\k_2^2)\mathbb{I}+f\big)^{-1}
 \big(i \del_- v  + \k_2^2 g^{-1} \del_- g\big)\Big) \ .
\end{split}
\end{equation}
It can be shown that this action is the non-Abelian T-dual of
\begin{equation*}
S = -{1\ov \pi} \int \text{d}^2\s\, \Tr\Big( \k_1^2 \tilde g^{-1} \del_+ \tilde g
\tilde g^{-1} \del_- \tilde g
+\k_2^2  (g^{-1}\del_+ g -\tilde g^{-1}\del_+ \tilde g)
(g^{-1}\del_- g -\tilde g^{-1}\del_- \tilde g)\Big)\ ,
\end{equation*}
with respect to  the global symmetry $\tilde g \mapsto \L \tilde g$, $\L\in G$.
{
Note that, if we define the new group element $\widetilde{\mathcal{G}}=g\tilde g^{-1}$ one may write the previous action as
\be
S=  -{1\ov \pi} \int \text{d}^2\s\, \Tr\Big(\k_1^2\, \tilde g^{-1} \del_+ \tilde g \tilde g^{-1} \del_- \tilde g +
\k_2^2\, \widetilde{\mathcal{G}}^{-1} \del_+ \widetilde{\mathcal{G}}
\widetilde{\mathcal{G}}^{-1} \del_- \widetilde{\mathcal{G}} \Big)\ ,
\ee
which is the sum of two independent PCM actions for a group $G$.}
The previous group element redefinition introduces interactions between them.

\no
Finally consider a limit in which only $\lambda_2$ tends to one, whereas $\lambda_1$ stays inactive. Then,
\eqref{dsh11} has to be modified as
\begin{equation*}
\lambda_2=1-{\k_2^2\ov k}\ ,\quad \mathcal{G}= \mathbb{I} + i {v\ov \sqrt{k}}\ ,\quad k\to \infty\ ,
\end{equation*}
in order for \eqref{rewr} to stay finite. In particular, this becomes
\be
S_{\k^2}(v,g)={1\ov 2\pi} {1+\lambda_1\ov 1-\lambda_1} \int \text{d}^2\s\, \Tr(\del_+ v\del_-v)-
{\k_2^2\ov \pi} \int \text{d}^2\s\, \Tr(g^{-1}\del_+ g
g^{-1}\del_- g)\ ,
\ee
representing $\dim G$ free bosons and a PCM model for a group $G$. This is consistent with the limit of
the $\beta$-functions for $\lambda_1$ and $\lambda_2$ (see, eqs. (2.6) and (2.7) in \cite{Georgiou:2017aei}). In this limit, the constant
$\lambda_1$ does not run since it can be absorbed into a redefinition of the $v$'s. Also the coupling constant $\k_2^2$
obeys the same RG flow equation appropriate for the PCM model and its non-Abelian T-dual, since these models are
canonically equivalent.

\no
It would be very interesting to explore physical applications in an AdS/CFT context
of this version of non-Abelian T-duality along the lines and developments
of \cite{Sfetsos:2010uq,Itsios:2012zv,Itsios:2013wd,Lozano:2016kum,Lozano:2016wrs,Itsios:2016ooc,Lozano:2017ole} (for a partial list of
works in this direction). Prototype examples this can be applied are the backgrounds
$\text{AdS}_3\times \text{S}^3 \times \text{S}^3 \times \text{S}^1$
and $\text{AdS}_5\times \text{S}^5$.

\section{Cyclic $\lambda$-deformations}
\label{cycliclambdasection}

In this section we construct a class of multi-parameter deformations of conformal field theories of the WZW type
Consider $n$ WZW models and $n$ PCMs for a group $G$, { defined with group elements $g_i$ and $\tilde g_i$, respectively.} We would like
to gauge the global symmetry
\begin{equation*}
g_i \mapsto \L_i^{-1}\, g_i\, \L_{i+1}\ ,\quad \tilde g_i \mapsto \L^{-1}_i \tilde g_i\ ,
\quad i=1,2,\dots , n,
\end{equation*}
with the periodicity condition $\L_{n+1}=\L_1$ implied.
We introduce gauge fields $A^{(i)}_{\pm}$ in the Lie-algebra of $G$ transforming as
\begin{equation}
A^{(i)}_{\pm} \mapsto \L_i^{-1} A^{(i)}_{\pm} \L_i - \L_i^{-1}\del_\pm \L_i\ ,\quad i=1,2,\dots ,n\ .
\end{equation}
In this way we have a periodic chain of interacting models each one of which separately is gauge anomalous by a term
independent of the group elements. The full model has no gauge anomaly since these cancel among
themselves (the chain may be open as long as it is infinite long). The details are quite similar to those for
the $n=2$ case \cite{Georgiou:2016urf}, so that we omit them here.

The choice $\tilde g_i =\mathbb{I}$, $i=1,2,\dots , n $ completely fixes the gauge and is consistent
with the equations of motion for the group elements $\tilde g_i$ of the PCMs which are automatically satisfied.
Then, the gauged fixed action becomes
\be
\label{action333}
\begin{split}
&
S_{k,\lambda_i}(\{g_i;A^{(i)}_{\pm}\}) = \sum_{i=1}^n S_k(g_i)
+ {k\ov \pi} \int \text{d}^2\s\sum_{i=1}^n \Tr\Big(
A^{(i)}_- \del_+ g_i g^{-1}_i - A^{(i+1)}_+ g_i^{-1} \del_- g_i
\\
&
\qq\qq \qq + A^{(i)}_- g_i A^{(i+1)}_+\, g_i^{-1} -A^{(i)}_+ \lambda_i^{-1} A^{(i)}_-\Big)\ ,
\end{split}
\ee
where the index $i$ is defined modulo $n$.
The equations of motion with respect to the $A^{(i)}_{\pm}$'s are given by
\begin{equation*}
\lambda_i^T D_i A^{(i+1)}_+ - A^{(i)}_+ =- i \lambda_i^T J^{(i)}_{+}\ ,\quad
\lambda_{i+1} D_i^T A^{(i)}_- - A^{(i+1)}_{-} = i \lambda_{i+1} J^{(i)}_{-}\ .
\end{equation*}
Solving them we find that
\begin{equation}
\begin{split}
A^{(1)}_{+}=i (\mathbb{I}- x_1 x_2\cdots x_n )^{-1} \sum_{i=1}^n x_1 x_2 \cdots x_{i-1} \lambda_i^T J^{(i)}_{+}\,,\quad
x_i=\lambda_i^T D_i\ .
\end{split}
\end{equation}
The rest can be obtained by cyclic permutations.
Plugging the latter into \eqref{action333} we find that the on-shell action reads
\begin{eqnarray}
\label{defactigenn}
&& S_{k,\lambda_i}(\{g_i\})=  {k\ov 12\pi}  \int  \Tr(g_1^{-1} \text{d}g_1)^3
+ {k\ov \pi} \int \text{d}^2\s\,{\rm Tr}\Bigg(\ha J^{(1)}_{+} D_1
{\mathbb{I} +x_1^T x_n^T x_{n-1}^T \cdots x_2^T\ov  \mathbb{I} - x_1^T x_n^T x_{n-1}^T \cdots x_2^T} J^{(1)}_{-} \nonumber
\\ &&
+ \sum_{i=2}^n J^{(i)}_{+} \lambda_i x_{i-1}^T\cdots x_2^T (\mathbb{I} -x_1^T x_n^T x_{n-1}^T \cdots x_2^T)^{-1} J^{(1)}_{-}\Bigg)
 \ +\ {\rm cyclic\ in}\ 1,2,\dots , n\ ,
\end{eqnarray}
where we have separated the Wess--Zumino term from the WZW model action.
For small values of the matrices we have that
\be
\label{fhh22}
S_{k,\lambda_i}(\{g_i\}) = \sum_{i=1}^n S_k(g_i)
+ {k\ov \pi} \sum_{i=1}^n \int \text{d}^2\s\, {\rm Tr}\left(J^{(i+1)}_+ \lambda_{i+1} J^{(i)}_-\right) + {\cal O}(\l^2)\ ,
\ee
representing $n$ distinct WZW models interacting by mutual current bilinears, for which \eqref{defactigenn} is the
all loop, in the $\lambda_i$'s, effective action.

We would like to stress that the $n=2$ is significantly different with
respect to higher $n$'s. Firstly,
the {\it non-perturbative} symmetry $\lambda_i\mapsto\lambda_i^{-1}$ and $k\mapsto-k$, is seemingly realized at a local level
for the group elements
only when $n=2$, see \eqref{symmdual} (also for $n=1$, see \eqref{duallli}). For higher values of $n$ the group
elements need to be transformed non-locally by using 
$\mathcal{J}^{(i)}_{\pm} \mapsto -\mathcal{J}^{(i+1)}_{\mp}$, with $n+1 \equiv 1$.
There are exceptions to this. In particular, if all $\lambda_i$ are
equal and isotropic, i.e. $\lambda_i=\l \mathbb{I}$, then this duality-type symmetry is
\be
k\mapsto -k \ ,\quad \l \mapsto {1\ov \l} \  ,
\quad
g_1\leftrightarrow g_2^{-1}\ , \quad g_n\leftrightarrow g_3^{-1}\ ,\quad g_{n-1} \leftrightarrow g_4^{-1}\ ,\quad {\rm etc}\ ,
\ee
that is the group elements are paired up as above. For odd $n$ one group element simply gets inverted.
Despite the fact that the symmetry can not be realized locally for the generic case it is still powerful enough
to constrain the $\b$-functions and current correlation functions of the cyclic model to have the same values
as those of the single $\lambda$-deformations.

\no
A second remark concerns the form of the action  \eqref{defactigenn} when one of the coupling matrices vanishes.
Consider this action for $n=2$ and $n=3$ when $\lambda_1=0$ while the other coupling matrices stay general
\ba
&&S_{k,0,\lambda_2}(g_1,g_2)=\sum_{i=1}^2S_k(g_i)+\frac{k}{\pi}\,\int\text{d}^2\s\, {\rm Tr}\left(J_+^{(2)}\lambda_2J_-^{(1)}\right)\,,\\
&&S_{k,0,\lambda_2,\lambda_3}(g_1,g_2,g_3)=\sum_{i=1}^3S_k(g_i)+\frac{k}{\pi}\,\int\text{d}^2\s\,{\rm Tr}\left(J_+^{(2)}\lambda_2J_-^{(1)}+
J_+^{(3)}\lambda_3J_-^{(2)}+J_+^{(3)}\lambda_3D_2^T\lambda_2J_-^{(1)}\right)\,.\nonumber
\ea
When $n=2$ the exact expression matches the approximate one in \eqref{fhh22}, while for $n=3$
the last term couples the three WZW models and it is quadratic in the $\lambda$'s.

\subsection{Algebra and Hamiltonian}

Here we provide the proof that the $\sigma$-model action \eqref{defactigenn}
is integrable for specific choices of the matrices $\lambda_i$, $i=1,2,\dots , n$.
In particular, we will show that it is integrable for all choices of the deformation
matrices $\lambda_i$ which, separately, give an integrable $\lambda$-deformed
model. These include the isotropic $\lambda$ for semi-simple group and symmetric coset, the anisotropic $SU(2)$ and the
$\lambda$-deformed Yang--Baxter model \cite{Sfetsos:2013wia,Sfetsos:2014lla,Sfetsos:2015nya,Hollowood:2014rla,Hollowood:2014qma}.

It is equivalent and more convenient to work with the gauged fixed action before integrating out the gauge fields.
Varying the gauged fixed action with respect $A^{(i)}_-$ and
$A^{(i+1)}_+$ we find the constraints
\be
\nabla_+ g_i\, g_i^{-1} =  (\lambda_i^{-T}-\mathbb{I}) A^{(i)}_+ \ ,\quad
g_i^{-1} \nabla_- g_i = - (\lambda_{i+1}^{-1}-\mathbb{I}) A^{(i+1)}_{-}\ ,
\label{dggd}
\ee
respectively. Varying with respect to $g_i$ we obtain  that
\be
\label{eqg1g2}
\nabla_-(\nabla_+ g_i g_i^{-1})= F_{+-}^{(i)}\ ,\quad \nabla_+( g_i^{-1}\nabla_- g_i)
= F_{+-}^{(i+1)}\ ,
\ee
which are in fact equivalent and where
$F_{+-}^{(i)}=\del_+ A^{(i)}_- - \del_- A^{(i)}_+ - [A^{(i)}_+,A^{(i)}_-]$.

Substituting \eqref{dggd} into \eqref{eqg1g2} we obtain after some algebra
that
\be
\begin{split}
\label{eomAinitial1}
&\del_+ A^{(i)}_- - \lambda_i^{-T} \del_- A^{(i)}_+ = [\lambda_i^{-T} A^{(i)}_+,A^{(i)}_-]\ ,
\\
& \lambda_i^{-1} \del_+ A^{(i)}_- -\del_-A^{(i)}_+=[A^{(i)}_+,\lambda_i^{-1}A^{(i)}_-]\ .
\end{split}
\ee
Hence the equations of motion split into $n$ identical sets which are seemingly decoupled
even though the $A^{(i)}_{\pm}$ depend on all group elements $g_i$ and coupling matrices $\l_i$, $i=1,2,\dots , n$.
Moreover, each set is the same one that one would have obtained had
{we} performed the corresponding analysis for the $\lambda$-deformed action \eqref{laact}.
Working along the lines of subsection \ref{doublyldeformed}; Eqns. \eqref{Affine}--\eqref{Hdensity0}
we find (for $n=2$ this was performed in detail in \cite{Georgiou:2016urf})
\be
\begin{split}
\label{Bowcockn}
&\{\mathcal{J}_{\pm}^{(i)a} , \mathcal{J}_{\pm}^{(i)b}\} =
\frac{2}{k}\,f_{abc} \mathcal{J}_{\pm}^{(i)c} \d_{\s\s'} \pm \frac{2}{k} \d_{ab}\, \d'_{\s\s'}\,,\\
&\mathcal{J}^{(i)}_{+} =\lambda_i^{-T} A^{(i)}_+ - A^{(i)}_-\ ,
\quad \mathcal{J}^{(i)}_{-} = \lambda_{i+1}^{-1} A^{(i+1)}_{-} - A^{(i+1)}_{+}\
\end{split}
\ee
and as a consequence
$
\{A^{(i)}_{\pm},A^{(j)}_{\pm}\}=0$, for $i\neq j$,
for all choices of signs and for generic coupling matrices $\lambda_{i}$.
Hence, all choices for matrices known to give rise to integrability
for the $\lambda$-deformed models provide integrable models here as well with independent conserved
changes.
The Hamiltonian density of the system in terms of $A^{(i)}_\pm$ and $\lambda_i$ is
\be
\label{Hdensitycyclic}
\boxed{
\mathcal{H}_{\text{cyclic}}=\frac{k}{4\pi}\sum_{i=1}^n{\rm Tr}\left(A_+^{(i)}\left(\lambda_i^{-1}\tilde h_i\lambda_i^{-T}\right)A_+^{(i)}+
A_-^{(i)}\left(\lambda_i^{-T} h_i\lambda_i^{-1}\right)A_-^{(i)}\right)}\,.
\ee
Using the above we generalize the result of subsection \ref{doublyldeformed}, that the cyclic 
$\lambda$-deformed {models} are
canonically equivalent to $n$ single $\lambda$-deformed $\sigma$-model.
The relations which define the canonical transformation are given by:
$
A^{(i)}_{\pm} =\tilde A^{(i)}_{\pm}\ ,\quad i=1,2,\dots, n\ ,
$
where the gauge fields without the tildes correspond to the cyclic deformed models and depend on
$(\lambda_1,\dots,\lambda_n;g_1,\dots,g_n)$, while those with tildes correspond to the canonically
equivalent sum of $n$ single $\lambda$-deformed models each one depending on $(\lambda_i;\tilde g_i)$.

\subsection*{RG flows and currents anomalous dimensions}

Similar to the case with $n=2$ considered in \cite{Georgiou:2017aei}, the expression \eqref{fhh22} can be used to argue
that the RG flow equations of the $n$ coupling matrices $\lambda_i$
for the cyclic model \eqref{defactigenn} as well as the currents anomalous dimensions are the same with
those obtained for the single $\lambda$-deformations model \cite{Itsios:2014lca,Sfetsos:2014jfa,Appadu:2015nfa}.
The  basic reason is that the various interaction terms have regular OPE among themselves so that correlations functions
involving currents factorize to those of $n$ single $\lambda$-deformed models.
This is also in agreement with the fact that the cyclic model is canonically equivalent to $n$ single $\lambda$-deformations.
Furthermore we mention without presenting any details that using the analysis performed in
\cite{Appadu:2015nfa,Georgiou:2017aei} we have explicitly checked the above claim
for the cases of $n$ isotropic couplings for general groups and symmetric spaces.

\subsection*{Acknowledgements}

K. Sfetsos would like to thank the
Physics Division, National Center for Theoretical Sciences of the National Tsing-Hua
University in Taiwan and the Centre de Physique Th\'eorique,
\'Ecole Polytechnique for hospitality and financial support during initial stages of this work.
G. Georgiou and K. Siampos acknowledge the Physics Department of the National 
and Kapodistrian University of Athens for hospitality. Part of this work was developed 
during HEP 2017: Recent Developments in High Energy Physics and Cosmology 
in April 2017 at the U. of Ioannina.

\appendix

\end{document}